# Riding the crest of the altmetrics wave: How librarians can help prepare faculty for the next generation of research impact metrics

Scott Lapinski, Heather Piwowar, Jason Priem

*Scott Lapinski is digital resources and services librarian at the Countway Library, Harvard Medical School, email: P_Lapinski@hms.harvard.edu, Heather Piwowar is a co-founder of ImpactStory and a postdoc at the National Evolutionary Synthesis through Duke University, email heather@impactstory.org, and Jason Priem is a co-founder of ImpactStory and PhD student at the School of Information and Library Science, UNC-Chapel Hill, email jason@impactstory.org*

Over the last decade, scholars have begun a great migration into online spaces, moving workflows and discussions to online platforms like Mendeley, blogs, Twitter, Facebook, and more. In these new spaces, once-invisible interactions like reading, saving, discussing, and recommending become visible. They leave traces. Observing these traces can inform new metrics of scholarly influence and impact -- so-called *altmetrics* [1].

These altmetrics are fast: data appears in days or weeks, instead of the years required by citations. More importantly, they are *diverse*, tracking impacts all across a quickly-changing scholarly communications landscape, populated by

- Diverse *products* beyond the article, including datasets, software, and blog posts,
- Diverse *platforms* beyond the traditional journal, like institutional repositories and online communities, and
- Diverse *audiences* beyond the academy, including practitioners, clinicians, and the general public.

University faculty, administration, librarians, and publishers alike are beginning to discuss how and where altmetrics can be useful towards evaluating a researcher's academic contribution [2]. As interest grows, libraries are in a unique position to help facilitate an informed dialog with the various constituencies that will intersect with altmetrics on campus, including both researchers (students and faculty) and the academic administrative office (faculty affairs, research & grants, promotion & tenure committees, and so on).

Librarians can provide this support in three main ways: informing emerging conversations with the latest research, supporting experimentation with emerging altmetrics tools, and engaging in early altmetrics education and outreach.

## Know the literature

Librarians can begin by familiarizing themselves with the current state of discussion around altmetrics.  Good places to start include a recent SPARC report [3], Galligan and Dyas-Correia's excellent overview [2], and the recent ASIS&T Bulletin special issue on altmetrics [4]. Librarians should also be familiar with the growing body of peer-reviewed research on altmetrics. An important concept from this literature is the idea of "impact flavors," a way to understand the distinctive patterns in the diverse impacts of individual products. A product featured in mainstream media stories, blogged about, and downloaded by the public, for instance, has a very different flavor of impact than one heavily saved and discussed by scholars, which is in turn different from one highly cited in research papers. Altmetrics can help researchers, funders, and administrators optimize for the mix of flavors that best fits their particular goals. [5]

Other noteworthy research has examined correlations between altmetrics and traditional citations finding that some altmetrics sources, particularly Mendeley, are significantly correlated with citation (around .5 in several studies) [5,6,7]. This same research shows that other sources, like Facebook bookmarks, correlated only slightly with citations; this suggests that they track different kinds of impacts. Other early touchstones include studies exploring the predictive potential of altmetrics [8], growing adoption of social media tools that inform altmetrics [9], and insights from article readership patterns [10].

## Know the tools

Altmetrics are in active use today: several tools allow scholars to collect and share the broad impact of their research portfolios.  In the same way a librarian would experiment with new features added to a once-familiar search interface just before the fall semester, librarians can play around with altmetrics tools to add them to their bibliographic instruction repertoire. Familiarity will enable a librarian to do easy demonstrations, discuss strengths and weaknesses, contribute to product development direction, and serve as a resource point for campus scholars and administration during the upcoming transition to web-native scholarship.

A great place to start experimenting is ImpactStory, a nonprofit web application created by two of this article's coauthors (JP and HP) and supported by the Alfred P. Sloan Foundation. Scholars upload their articles, datasets, software repositories, and other products to ImpactStory using Google Scholar, ORCID, or DOI lists.  ImpactStory then gathers and reports both altmetrics and traditional citations for each product.  As shown below, metrics are displayed as percentiles relative to similar products; data can be exported for further analysis. ImpactStory is built on open-source code, offers open data, and is free to use (see http://impactstory.org).

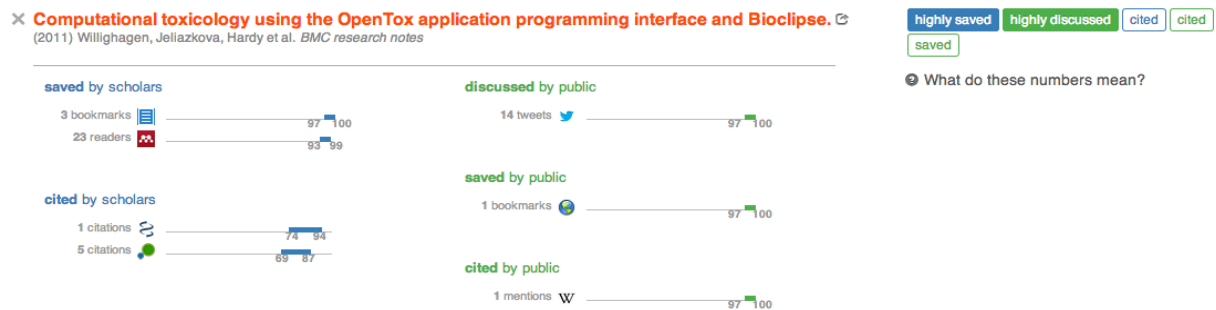

Figure 1: ImpactStory report for a BioMed Central article

PlumX is another tool that provides impact profiles for scholars. Like ImpactStory, PlumX is a web application that displays altmetrics on a wide range of scholarly products. PlumX is available to scholars upon university-wide subscription; users can experiment with a free demo version (see http://www.plumanalytics.com).

## Integrate altmetrics into library outreach and education

Establishing a strong familiarity with the altmetrics tools will allow librarians to enhance standard bibliographic instruction with this added perspective. As these opportunities for outreach and engagement present themselves, librarians will find that even a brief demo of tools like ImpactStory and its ability to pull together usage data from a variety of resources hosting a scholar's work, will stimulate interest among content producers and users alike.

Requests for using specific research databases and understanding publishing choices along with preparing bibliographic formatting to an author's manuscript are common requests at the library. As students and faculty engage with the library during any of these activities, either via a drop-in research consultations, or invitations to speak at student orientations and faculty meetings, these opportunities allow us to touch on altmetrics in demonstrating where it will intersect with their published research. Showing article level metrics, as in the PLoS journals, or download statistics from PubMedCentral and other institutional repositories allows for a quick introduction to where they (as "content producer") may consider participating. Depending on the options available on one's campus, it could also be an opportunity to highlight the benefits of participating in open access venues, and discuss "impact" as something more closely tied to an individual's scholarship rather than "a number" set exclusively to a journal title. The conversation can also segue into a useful discussion about the limitations imposed by placing too much reliance on the more familiar Journal Impact Factor (JIF) or h-index (an author oriented metric based on a calculation between the number of articles an author has published and citations received for each of those same articles).

Whether or not traditional measures like JIF or h-index are being "encouraged" by various faculty or departments on campus to measure the impact of a scholar's article output, by introducing these same faculty and departments to altmetrics, the library can play an important role. It may be the case where a researcher or a faculty affairs department will request the library to provide assistance and instruction on calculating an h-index within various databases (Web of Science, SCOPUS, etc.). Integrating altmetrics into these instruction sessions is in the same spirit as a library providing a researcher with several additional choices in considering primary resources on any research project. We need to make researchers aware of the choices that are available to them in evaluating the impact of scholarship, and the relevant research, helping them make informed choices.

Libraries should take advantage of opportunities to demonstrate ImpactStory to multiple constituency groups and share visual information on how these metrics have been integrated into a researcher's profile. If there has not yet been an opportunity for the faculty member to visualize usage data from a conference presentation on SlideShare, or a video interview that may have been posted on Vimeo, here is a chance to, if not peak interest, at least stimulate awareness to the possibilities. The added benefit of sharing these examples with researchers on campus may extend beyond just introduction to altmetrics, and provide a window into the online communities that are sharing scholarship in ways in which the researcher had not yet considered participating in.

## Conclusion

Traditional impact measures, most commonly the Journal Impact Factor and the h-index, continue to be the source of much debate, and over the years have provoked many suggestions for ways in which their interpretation (or algorithms) could be improved. Altmetrics is not a complete answer to whatever shortcomings are inherent within these traditional impact measures. However, altmetrics do allow assessment directly at the product level, rather than the publication. Moreover, they cover the growing diversity in scholarly products, platforms, and people.

Of course, early excitement in altmetrics' potential must be tempered by appropriate caution; research into the validity and reliability of altmetrics is still in its infancy. However, as we transition from a paper-native to a web-native scholarly communication system [11], these new metrics are likely to grow in importance. As they do, librarians are well positioned to inform and support researchers and decision makers in their use.